\documentclass[sigconf,natbib=true,screen=true]{acmart}

\acmSubmissionID{3515}

\usepackage{graphicx} %
\usepackage{geometry}
\usepackage{array}
\usepackage{enumitem}
\usepackage[acronym]{glossaries}
\makeglossaries
\usepackage{multirow, makecell}
\usepackage{tikz}
\usepackage{adjustbox}
\usepackage{booktabs}
\usepackage{diagbox} 
\usetikzlibrary{arrows.meta, positioning, shapes.geometric}
\usepackage{tabularx}
\usepackage{booktabs}
\usepackage{threeparttable}
\newcolumntype{Y}{>{\centering\arraybackslash}X} 
\usepackage[skip=2pt]{caption}
\usepackage{mleftright}
\usepackage{hyperref}

\newcommand{\headernodot}[1]{\vspace{1mm}\noindent\textbf{#1}}
\newcommand{\heading}[1]{\headernodot{#1.}}

\allowdisplaybreaks

\author{Jingwei Kang}
\authornote{Both authors contributed equally to this research.}
\affiliation{%
  \institution{University of Amsterdam}
  \city{Amsterdam}
  \country{The~Netherlands}
}
\email{j.kang@uva.nl}
\orcid{0009-0003-9283-4060}

\author{Santiago de Leon-Martinez}
\authornotemark[1]
\affiliation{%
  \institution{Brno University of Technology}
  \city{Brno}
  \country{Czechia}}
\additionalaffiliation{%
  \institution{Kempelen Institute of Intelligent Technologies}
  \city{Bratislava}
  \country{Slovakia}
}
\email{santiago.deleon@kinit.sk}
\orcid{0000-0002-2109-9420}

\author{Maarten de Rijke}
\orcid{0000-0002-1086-0202}
\affiliation{%
  \institution{University of Amsterdam}
  \city{Amsterdam}
  \country{The~Netherlands}
}
\email{m.derijke@uva.nl}

\author{Harrie Oosterhuis}
\orcid{0000-0002-0458-9233}
\affiliation{%
  \institution{University of Amsterdam}
  \city{Amsterdam}
  \country{The~Netherlands}
}
\email{h.r.oosterhuis@uva.nl}

\copyrightyear{2026}
\acmYear{2026}
\setcopyright{cc}
\setcctype{by}
\acmConference[CIKM '26]{Proceedings of the 35th ACM International Conference on Information and Knowledge Management}{November 07--11, 2026}{Rome, Italy}
\acmBooktitle{Proceedings of the 35th ACM International Conference on Information and Knowledge Management (CIKM '26), November 07--11, 2026, Rome, Italy}
\acmDOI{10.1145/3799682.3839894}
\acmISBN{979-8-4007-2539-5/2026/11}



\title[Revisiting N2DCG: An Empirically Grounded Reformulation of Carousel Recommendation Evaluation]{Revisiting N2DCG: An Empirically Grounded Reformulation\\ of Carousel Recommendation Evaluation}

\begin{document}

\begin{abstract}
Carousel interfaces have been widely used in video and music streaming services, yet it remains unclear how to properly evaluate recommender systems in these two-dimensional layouts. 
N2DCG has been proposed to address this gap by adapting NDCG to carousel-based recommendation, but it relies on unverified assumptions borrowed from the single-list web-search setting that do not transfer well to two-dimensional carousel layouts.

We identify two substantial limitations of N2DCG: its ideal ranking, used for normalization, violates carousel constraints, and its discount function does not reflect user browsing behavior observed in empirical data.
To address both limitations, we propose a reformulation of N2DCG that normalizes appropriately by respecting constraints and uses an empirically grounded discount function.
We validate the proposed metric, showing that it better reflects users' empirical behavior on real-world eye-tracking data and better predicts the comparison results of carousel layouts simulated based on empirical examination patterns.
\end{abstract}

\maketitle

\glsresetall

\section{Introduction: New Insights in Carousels}
Carousel interfaces have become ubiquitous in streaming media services like Netflix~\citep{10.1145/2843948, wang2026genpage}. They encourage user exploration by containing multiple lists or rows called \textit{carousels}, each horizontally swipeable and organized around a single defining theme.
Despite their widespread use, carousels have only recently gained attention in recommender systems research, with work on user click modeling \cite{10.1145/3643709, deleonmartinez2026latentobservablepositionbasedclick} and on evaluation \cite{ferraridacrema2022offline,10.1145/3450614.3461680}.
However, a recent eye-tracking study \cite{10.1145/3726302.3730301} and two follow-up analyses \cite{10.1145/3742413.3789166,kang2026followingeyetrackingevidenceestablished} reveal a behavioral pattern that prior work does not account for.
User examination in carousels does not follow a single global F-pattern \cite{10.1145/3726302.3730301, 10.1145/3742413.3789166,kang2026followingeyetrackingevidenceestablished}: the initial page shows a dual-focus F-pattern (with attention peaks at both the top-left and top-right corners), and after swiping this shifts to a mirrored F-pattern, with horizontal decay running right-to-left rather than left-to-right. 
This behavior is attributed to users maintaining their focus on the rightmost position after swiping.
However, N2DCG \cite{ferraridacrema2022offline,10.1145/3450614.3461680}, the only evaluation metric for carousels, does not account for this behavior. 
Our work addresses this gap by re-examining and reformulating N2DCG.

N2DCG is a two-dimensional extension of Normalized Discounted Cumulative Gain (NDCG).
NDCG evaluates a ranked list by combining item relevance with position discount.
The (unnormalized) DCG of a list $y$ with $J$ items is defined as \cite{10.1145/345508.345545,10.1145/582415.582418,10.1145/1102351.1102363}:
\begin{equation}
    DCG(y) = \sum_{j=1}^{J} g(y_j) \cdot d(j) = \sum_{j=1}^{J} \frac{g(y_j)}{\log_2(j+1)},
\end{equation}
where $y_j$ denotes the item at rank $j$, $g(\cdot)$ represents the gain function mapping an item to its relevance score, and $d(j)$ is the position discount factor, commonly represented by a logarithm. To ensure comparability across different queries, DCG is subsequently normalized by the Ideal DCG (IDCG):
\begin{equation}
IDCG = \max_{y^{*} \in \mathcal{Y}} DCG(y^{*}),
\quad
    NDCG(y) = \frac{DCG(y)}{IDCG},
\end{equation}
where $\mathcal{Y}$ denotes the set of all possible permutations of candidate items for the given query;
IDCG represents the maximum possible DCG score for a ranking of the given items.

Standard NDCG assumes a 1D ranked list, whereas carousel interfaces present items in a 2D layout. 
To account for row and column positions, 2D Discounted Cumulative Gain (2DCG) extends DCG to an $I \times J$ grid \cite{ferraridacrema2022offline,10.1145/3450614.3461680}. Let $y_{i,j}$ denote the item located at the $i$-th row and $j$-th column in a 2D arrangement $y$:
\begin{equation}
    2DCG(y) = \sum_{i=1}^{I} \sum_{j=1}^{J} g(y_{i,j}) \cdot d(i,j),
\end{equation}
where $I$ denotes the number of rows (e.g., carousels) and $J$ the number of items in each row. 
The gain term $g(\cdot)$ remains the same as in DCG, whilst the 2D discount term $d(i,j)$ models the combined effect of visual attention and interaction effort. 
The design of $d(i,j)$ is the key modeling choice for carousels. \citet{ferraridacrema2022offline} propose two formulations:
\begin{equation}
\label{naive F action}
\mbox{}\hspace*{-2.5mm}
d_t(i,j)=\frac{1}{\log_2(\alpha i+\beta j)},\;
d_a(i,j)=\frac{1}{\log_2(\alpha i+\beta j+\gamma n_h+\lambda n_v)}.
\hspace*{-2.5mm}\mbox{}
\end{equation}
Here, $\alpha$ and $\beta$ penalize lower row positions $i$ and farther-right column positions $j$, respectively, reflecting F-pattern behavior.
For $d_a$, $\gamma$ and $\lambda$ additionally penalize the number of horizontal swipes $n_{h}$ and vertical scrolls $n_{v}$ required to reveal an item.

Similar to NDCG, 2DCG is normalized by the Ideal 2DCG (I2DCG) to obtain a comparable metric bounded between 0 and 1:
\begin{equation}
     I2DCG = \max_{y^{*} \in \mathcal{Y}_{2D}} 2DCG(y^{*}), \quad N2DCG(y) = \frac{2DCG(y)}{I2DCG},
\end{equation}
where I2DCG represents the maximum possible 2DCG score achieved by mapping the most relevant items to grid coordinates with the highest position discount factors.
Here, $\mathcal{Y}_{2D}$ denotes the set of all possible 2D permutations of candidate items.
\citet{ferraridacrema2022offline} express this mathematically as: the ideal arrangement $y^{*}$ ensures $g(y^{*}_{i,j}) \ge g(y^{*}_{i',j'})$ for any pair of positions where $d(i,j) > d(i',j')$ .

Building on this foundation, we address the limitations of N2DCG through the following contributions:
\begin{itemize}[leftmargin=*]
    \item We reformulate I2DCG to account for categorical constraints necessary for carousel interfaces and their evaluation.
    \item We identify and correct a failure of the original N2DCG discount factor to model mirrored F-pattern examination behavior.
    \item We design and evaluate four novel position-discount formulations; among these, the row-page discount aligns most closely with empirical examination frequencies and most reliably predicts user preferences between simulated carousel layouts.
\end{itemize}

\section{Limitations of the Original N2DCG}
\subsection{Ignoring Categorical Constraints}
The original N2DCG formulations ignore a key property of carousel interfaces: each carousel is, by design, organized around a single category or theme, and all items within it are drawn from that category.
I2DCG, as defined above, treats all $I \times J$ positions as a single global pool and assigns items solely to maximize 2DCG, with no awareness of these categorical constraints.
The resulting I2DCG values come from layouts that no valid carousel could realize, and dividing by them lowers the N2DCG scores of valid layouts.

Table~\ref{tab:n2dcg-example} illustrates this flaw using an example 2$\times$3 carousel. 
When relevant items belong to different categories, the unconstrained I2DCG  
simply places all of them into the top positions, producing an invalid layout that no carousel interface could display. 
In the example, the optimal valid layout reaches a 2DCG of 2.2, while the unconstrained optimal is 2.7.
As a result, the valid layouts can only reach a maximum N2DCG of 0.81.
This causes inconsistencies in evaluation as the actual upper bound of N2DCG is lower than 1.

\begin{table}[t]
\centering
\caption{A 2$\times$3 carousel interface with row discounts $(1.0, 0.9, 0.8)$
and $(0.3, 0.2, 0.1)$. Candidates span two categories,
$A=\{a_1,a_2,a_3\}$ and $B=\{b_1,b_2,b_3\}$. Relevance is binary:
$a_1, a_2, b_1$ are relevant, $a_3, b_2,b_3$ are non-relevant.
The unconstrained ideal mixes categories within rows, inflating the
N2DCG denominator.}
\label{tab:n2dcg-example}
\begin{tabular}{@{}l l r@{}}
\toprule
\textbf{Case} & \textbf{Layout} & \textbf{2DCG} \\
\midrule
Optimal valid layout
& $\begin{pmatrix} a_1 & a_2 & a_3 \\ b_1 & b_2 & b_3 \end{pmatrix}$
& $2.2$ \\
\addlinespace[4pt]
Optimal unconstrained layout
& $\begin{pmatrix} a_1 & a_2 & b_1 \\ a_3 & b_2 & b_3 \end{pmatrix}$
& $2.7$ \\
\midrule
\multicolumn{2}{@{}l}{Maximum N2DCG among valid layouts ($2.2/2.7$)} & $0.81$ \\
\bottomrule
\end{tabular}
\end{table}

\subsection{Mismodeling Mirrored F-Pattern Behavior}
A position discount function does not need to exactly match empirical examination frequencies. 
For evaluation, it is often sufficient for the discount to preserve the correct qualitative trend of user attention. 
However, even this weaker requirement is not met by the original N2DCG discount.

\begin{figure*}[t]
  \centering
  \includegraphics[width=\linewidth]{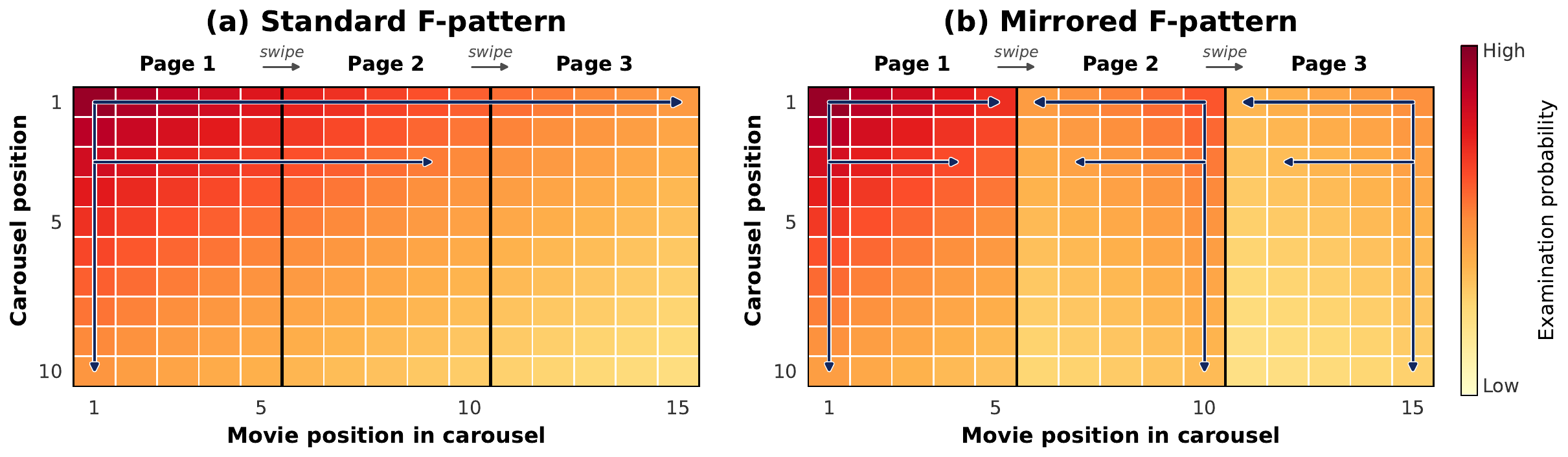}
\caption{Schematic comparison of the assumed~(a) and observed~(b) attention patterns in a carousel interface. Darker cells indicate higher examination probability; arrows indicate the direction of attention decay.}
  \label{fig:pattern-comparison}
\end{figure*}

In real carousel interfaces, users often navigate within a carousel through horizontal swiping, which creates page boundaries within each carousel. 
Eye-tracking evidence suggests that after users swipe to a new page, their attention may reset toward the right side of the newly visible page, so that on later pages attention decreases from right to left as a mirrored F-pattern \citep{10.1145/3726302.3730301,kang2026followingeyetrackingevidenceestablished}, as illustrated in Figure~\ref{fig:pattern-comparison}(b).
Because the N2DCG discount monotonically penalizes larger column positions, it instead predicts attention that keeps decreasing from left to right across the entire carousel (Figure~\ref{fig:pattern-comparison}(a)), the opposite of the observed trend on the second and third pages.
Thus, the original position discount captures the wrong direction of attention change after horizontal swiping.

\section{Proposal: Category-Aware N2DCG}
We propose a category-aware variant of N2DCG in which the ideal layout is
constrained to respect the category structure of the carousel interface.
Let $\mathcal{Y}_{2D}^{\mathrm{cat}} \subset \mathcal{Y}_{2D}$ denote the
subset of 2D layouts in which all items in a row belong to the same category:
\begin{equation}
\mathcal{Y}_{2D}^{\mathrm{cat}} = \left\{
y \in \mathcal{Y}_{2D}
\;\middle|\;
\forall\, i,j,j',\; \operatorname{cat}(y_{i,j}) = \operatorname{cat}(y_{i,j'})
\right\}, 
\end{equation}
where $\operatorname{cat}(\cdot)$ denotes the category of an item.
Since each category corresponds to exactly one carousel, $\mathcal{Y}_{2D}^{\mathrm{cat}}$ consists precisely of the layouts obtained
by (i) assigning category-specific carousels to display rows and (ii) ranking items within each carousel; items from different categories never share a row.
The category-aware $\mathrm{I2DCG}_{\mathrm{cat}}$ is the maximum 2DCG over this restricted space:
\begin{equation}
\mathrm{I2DCG}_{\mathrm{cat}} = \max_{y^{*} \in \mathcal{Y}_{2D}^{\mathrm{cat}}} \mathrm{2DCG}(y^{*}),
\end{equation}
and the category-aware $\mathrm{N2DCG}_{\mathrm{cat}}$ is defined as 
$\frac{\mathrm{2DCG}}{\mathrm{I2DCG}_{\mathrm{cat}}}$.

In our setting, computing $\mathrm{I2DCG}_{\mathrm{cat}}$ reduces to assigning category-specific carousels to display rows. 
Given a category-specific carousel and a display row, the optimal within-carousel ranking is obtained by placing items in decreasing relevance order onto that row's decreasing-discount positions. 
This yields a score for each carousel--row pair. 
The remaining problem is to map category-specific carousels to display rows to maximize the total 2DCG, which can be solved using assignment algorithms such as the Hungarian algorithm~\citep{kuhn-1955-hungarian}.

For clarity, this computation is only used for normalization during evaluation.
We are not proposing to use the Hungarian algorithm in production to create recommendations for actual users.

\section{Proposal: Mirrored F-Pattern N2DCG}
To capture both mirrored F-pattern attention and the interaction effort of accessing carousel items, we propose four formulations of the discount function $d(i,j)$ at row $i$ and column $j$.
We first modify the position discount to reflect attention shift after a swipe, and then explore three ways to incorporate user action penalties into the base discount.

\subsection{Mirrored F-Pattern Positional Discount}
The original position discount $d_t$ in Eq.~\ref{naive F action} models F-pattern position bias through a two-dimensional logarithmic decay.
To account for attention shift after horizontal swipes, we modify this decay:
\begin{equation}
    d_\text{mir}(i,j) = \frac{1}{\log_{2}(\alpha i + \beta \tilde{j})}
\end{equation}
Let $\delta$ denote the number of items visible per page within a carousel. After a swipe, a new page of $\delta$ items is displayed, and the user's attention is naturally drawn first to the newly displayed items on the right side, and subsequently shifts toward the left.
We capture this by replacing the original column index $j$ with a re-indexed position $\tilde{j}$:
\begin{equation}
\tilde{j} =
\begin{cases}
j, & j \le \delta, \\[4pt]
\delta\left\lfloor \dfrac{j-1}{\delta} \right\rfloor + \left(\delta - ((j-1)\bmod \delta)\right), & j > \delta.
\end{cases}
\end{equation}
On the first page ($j \le \delta$) the indexing is unchanged; on each subsequent page, the order within the page is mirrored, so the rightmost position carries the smallest $\tilde{j}$ and therefore the highest discount.

\subsection{Incorporating User Action Penalties}
Accessing items that are not currently visible requires additional swiping. Let $n_h$ and $n_v$ denote the number of horizontal and vertical swipes required to reveal the item at position $(i,j)$. 
We compare three strategies for penalizing these actions.

\heading{Additive Swipe Penalties} Following $d_a$ in Eq.~\ref{naive F action}, the swipe cost is added inside the logarithmic denominator with weight parameters $\gamma$ and $\lambda$:
\begin{equation}
    d_\text{add}(i,j) = \frac{1}{\log_{2}(\alpha i + \beta \tilde{j} + \gamma n_h + \lambda n_v)}.
\end{equation}

\heading{Multiplicative Swipe Penalties}
The swipe cost is instead applied multiplicatively. 
Let $\eta, \theta \in (0, 1)$ denote the penalty factors applied per horizontal and vertical swipe, respectively:
\begin{equation}
    d_\text{mul}(i,j) = \frac{1}{\log_{2}(\alpha i + \beta \tilde{j})} \cdot \eta^{n_h} \cdot \theta^{n_v}.
\end{equation}

\heading{Row-Page Discount} Instead of penalizing every swipe, this variant only penalizes items that require at least one horizontal swipe, combined with geometric decay across rows. Let $\mu \in (0,1)$ be the page penalty and let $\nu \in (0,1)$ be the row decay factor:
\begin{equation}
    d_\text{RPD}(i,j) = \frac{1}{\log_{2}(\alpha i + \beta \tilde{j})} \cdot \mu^{\mathbf{1}[j > \delta]} \cdot \nu^{i - 1},
\end{equation}
where $\mathbf{1}[j > \delta]$ equals 1 when the item is not on the first page and 0 otherwise.
Intuitively, this formulation treats the \textit{first} horizontal swipe as costly but does not penalize subsequent pages further, and also applies a multiplicative decay for each carousel row a user must enter to reach the item.

\section{Experiments }
This section details our experiments using the RecGaze eye-tracking dataset to estimate the parameters of our proposed discount functions and systematically validates the robustness of the reformulated N2DCG metric through simulated carousel layouts. Our code is available at \url{https://github.com/jkang98/carousel-metric}.

\begin{figure*}[t]
  \centering
  \includegraphics[width=\linewidth]{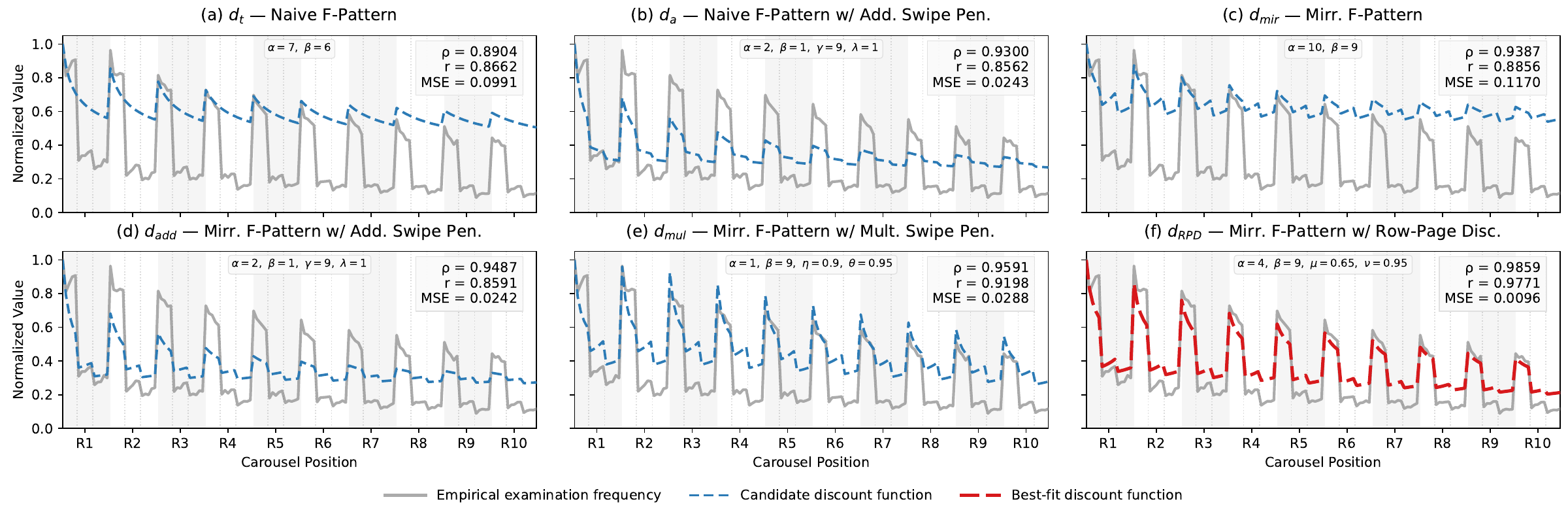}
\caption{Fitted parameters and test-set performance ($\rho$, $r$, MSE) of each discount function are reported. 
Both the empirical examination frequencies and the computed discounts are max-normalized to $[0, 1]$ before plotting and MSE evaluation, so the comparison reflects how well each formulation captures the \emph{relative} decay of user attention rather than its absolute magnitude.}
  \label{fig:comparison}
\end{figure*}

\heading{RecGaze Dataset} To date, RecGaze~\citep{10.1145/3726302.3730301} is the \emph{first and only} publicly available eye-tracking dataset designed for carousel recommendation.
It provides a unique empirical basis for modeling user examination behavior in carousels. 
We focus on the \emph{free-browsing task} of RecGaze because it best reflects natural exploration behavior in carousel interfaces. 
In this task, each screen contains ten genre-based carousels, with fifteen movies in each carousel divided into three pages of five movies, of which the first page is initially presented and the last two can be seen by horizontally swiping once or twice, respectively. The first three carousels are initially visible with the remaining requiring vertical scrolling.
This layout allows us to model both vertical carousel positions and horizontal within-carousel positions.

\heading{Experimental Setup} We use fixations as an observable proxy for item examination. 
Following the RecGaze preprocessing, an item is considered examined if at least one valid fixation falls within its area of interest. 
From the dataset, we estimate the examination probability at position $(i, j)$ with the observed frequency of examination: %
\begin{equation}
P\!\left(E_{i,j}=1\right) \approx
\hat{P}\!\left(E_{i,j}=1\right) \coloneq \frac{N(E_{i,j}=1)}{N_{\text{total}}},
\label{eq:exam-prob}
\end{equation}
where $N(E_{i,j}=1)$ denotes the number of screens in which the item displayed at position $(i,j)$ was examined and $N_{\text{total}}$ the total number of screens.
These empirical frequencies serve two purposes: they are the targets for fitting the parameters of each candidate discount function, and they provide the examination-based ground truth position weights in our simulation study below.

To compare different design choices for the discount function, we perform a train/test split based on the location of the participants. 
The training set is used to estimate the parameters of the discount function, while the test set is used to evaluate whether the learned function generalizes to a different user group. 
In our experiment, we use data from 61 participants in Bratislava, Slovakia for training and 26 participants in Amsterdam, the Netherlands for testing; after removing screens with accidental clicks, this yields 1665 train and 710 test screens, each containing one clicked item.

\heading{Discount Function Parameter Estimation}
We constrain the parameter search space based on the interpretation of each discount function. 
Across all variants, $\alpha$ and $\beta$ control the weights of the vertical carousel position and the horizontal item position inside the logarithm.
For additive action-based variants, $\gamma$ and $\lambda$ weight the numbers of horizontal swipes and vertical scrolls. 
Since these parameters weight integer-valued positions or actions, we restrict them to positive integers. 
In the multiplicative action-based variant, $\eta,\theta$ control horizontal/vertical action decay, and $\mu,\nu$ control page/row decay; all are constrained to $(0,1)$ as they directly scale the discount value.
To determine the optimal parameters, we conduct a grid search, varying $\alpha$, $\beta$, $\gamma$, and $\lambda$ over $[1, 10]$ in steps of $1$, and $\eta$, $\theta$, $\mu$, and $\nu$ over $(0, 1)$ in steps of $0.05$.

For each discount variant, we fit the optimal parameters on the training set by maximizing the Spearman's rank correlation coefficient ($\rho$). 
The goal is to identify the parameter setting that accurately preserves the ordinal ranking of the empirical examination frequencies observed at each position, ensuring that positions with higher expected attention correctly receive larger discount.

\heading{Discount Fit to Examination Data} Using the learned parameters, we evaluate each discount function against the empirical examination frequencies observed in the held-out test set.
Figure~\ref{fig:comparison} shows the results for different discount functions.
The gray curve represents the empirical examination frequency from eye-tracking data, and the dashed curves denote the fitted discount functions. 
Our best-performing formulation $d_{RPD}$ is highlighted in red.

The two naive F-pattern-based functions, $d_t$ and $d_a$, roughly capture the overall decreasing trend, but fail to reflect the repeated drops and rises caused by pagination; adding user-action information, $d_a$, improves the overall fit, but the pagination problem persists. 
The four mirrored F-pattern-based functions fit the empirical curve better, especially in capturing the repeated pagination pattern.
Within this family, implementing user actions multiplicatively ($d_{mul}$ and $d_{RPD}$) produces closer match. 
The best result is achieved by combining the mirrored  F-pattern with the row-page discount $d_{RPD}$, which yields the highest  Spearman's $\rho$ and Pearson's $r$ correlations and the lowest MSE after max normalization.
Accordingly, the next experiment compares the original and reformulated 2DCG using $d_a$ and $d_{RPD}$, respectively.

\heading{Metric Evaluation on Simulated Carousels} To systematically evaluate the reliability of our reformulated metric, we conduct a simulation experiment with 20,000 independent trials. 
Each trial samples a set of category-specific carousels: each carousel corresponds to a category and contains candidate items with fixed relevance labels.
We then construct two complete $10 \times 15$ carousel layouts, denoted as $A$ and $B$, by independently permuting the vertical order of the carousels and the horizontal order of items within each carousel. 
Therefore, the only difference between the two layouts is how carousels and items are arranged.
This experimental design evaluates only the position discount functions: because both layouts contain the same candidate items, any ideal normalizer would be identical for $A$ and $B$ and cancel in the pairwise comparison.
The category-aware $\text{I2DCG}_{\text{cat}}$ concerns the definition of the ideal layout rather than the modeling of user attention, and is therefore not the subject of this validation.

To obtain an empirical reference, we compute examination-based ground-truth scores, $N2DCG_A^{\mathrm{exam}}$ and $N2DCG_B^{\mathrm{exam}}$, as the category-aware $\mathrm{N2DCG}_{\mathrm{cat}}$ with its position discounts replaced by the empirical examination frequencies $\hat{P}(E_{i,j}=1)$ (Eq.~\ref{eq:exam-prob}), estimated on the held-out test set.
The resulting score reflects the expected relevance that users actually examine.
We then measure the difference between the two layouts as $\Delta = \left|N2DCG_A^{\mathrm{exam}} - N2DCG_B^{\mathrm{exam}}\right|$.
A near-zero $\Delta$ indicates that the two layouts are difficult to distinguish, while a larger $\Delta$ means that one layout is clearly better than the other in terms of actual user browsing behavior.

Finally, we score both layouts using the unnormalized 2DCG with the original discount $d_a$ and with our reformulated discount $d_{RPD}$.
A metric is considered correct in a trial if its pairwise preference matches the preference given by the empirical ground truth. 
By comparing the accuracy under different $\Delta$ thresholds, we can evaluate the robustness of each metric.
Table~\ref{tab:accuracy_comparison} reports accuracy under both binary and graded relevance (1--5 scale) settings.
Different thresholds are used for the ground-truth difference $\Delta$.
Here, \textbf{\textit{Corr.}} denotes the percentage of trials in which the original 2DCG makes an incorrect pairwise decision, while the reformulated 2DCG makes the correct one.
\textbf{\textit{Both Err.}} denotes the percentage of trials in which both metrics make incorrect decisions.
The results show that the reformulated 2DCG consistently achieves higher accuracy than the original metric. 
When the ground-truth difference is very obvious ($\Delta \geq 0.10$), the original 2DCG still misjudges $1.3\%$ (binary) / $1.8\%$ (graded) of trials.

\begin{table}[t]
\centering
\caption{Accuracy comparison between the original and reformulated 2DCG under different ground-truth differences.}
\label{tab:accuracy_comparison}
\resizebox{\columnwidth}{!}{%
\begin{tabular}{l c rr rr}
\toprule
\textbf{Setting} & $\textbf{Diff. $\Delta$} \ge$ 
& \textbf{Orig. Acc.} 
& \textbf{Our Acc.}
& \textbf{Corr.} 
& \textbf{Both Err.}\\
\midrule
\multirow{5}{*}{\textbf{Binary}} 
& 0.00 & 82.9\% & \textbf{93.2\%} & 11.4\% & 5.7\% \\
& 0.01 & 85.7\% & \textbf{96.4\%} & 11.0\% & 3.2\% \\
& 0.02 & 88.4\% & \textbf{98.3\%} & 10.0\% & 1.5\% \\
& 0.05 & 94.8\% & \textbf{99.9\%} & 5.2\% & 0.1\% \\
& 0.10 & 98.7\% & \textbf{100.0\%} & 1.3\% & 0.0\% \\
\midrule
\multirow{5}{*}{\textbf{Graded}} 
& 0.00 & 84.9\% & \textbf{93.9\%} & 10.4\% & 4.8\% \\
& 0.01 & 87.4\% & \textbf{96.7\%} & 9.9\% & 2.7\% \\
& 0.02 & 89.4\% & \textbf{98.2\%} & 9.1\% & 1.5\% \\
& 0.05 & 94.3\% & \textbf{99.8\%} & 5.5\% & 0.2\% \\
& 0.10 & 98.2\% & \textbf{100.0\%} & 1.8\% & 0.0\% \\
\bottomrule
\end{tabular}%
}
\end{table}

\heading{Lessons Learned}
Experiments show the row-page discount $d_{RPD}$ best captures user attention and preferences in carousel interfaces.
Thus, our reformulated 2DCG with our best found parameters is:
\begin{equation}
    2DCG_\text{ref}(y)
    = \sum_{i=1}^{I} 
    \sum_{j=1}^{J}
    \frac{g(y_{i,j})}
    {\log_{2}(4 i + 9 \tilde{j})}
    \cdot 0.65^{\mathbf{1}[j > \delta]}
    \cdot 0.95^{i - 1}
\end{equation}
We reformulate $\text{I2DCG}_{\text{cat}}$ and $\text{N2DCG}_{\text{cat}}$ correspondingly.

\section{Conclusion}
We identified two limitations of N2DCG---the neglect of categorical constraints and the misalignment with actual browsing behavior in carousel interfaces---and proposed a reformulation addressing both.
Validation using empirical examination frequencies and simulated pairwise comparisons demonstrates the effectiveness of our reformulation.
This suggests that our reformulated N2DCG is a better choice to evaluate carousel recommendation.
Future work should extend this metric-based evaluation into behavior-aware simulator-based evaluation, enabling carousel policies to be evaluated under explicit models of user browsing and choice behavior.

\begin{acks}
This work is supported by the Dutch Research Council (NWO) under grants
\href{https://www.nwo.nl/en/projects/viveni222269}{VI.Veni.222.269},
\href{https://www.nwo.nl/en/projects/024004022}{024.004.022},
\href{https://www.nwo.nl/en/projects/nwa138920183}{NWA.1389.20.183}, and
\href{https://www.nwo.nl/en/projects/kich3ltp20006}{KICH3.LTP.20.006};
by the European Union under grant agreement No.\
\href{https://doi.org/10.3030/101201510}{101201510} (UNITE);
and by the Eyes4ICU project, funded by the European Union under the Horizon
Europe Marie Sk\l{}odowska-Curie Actions, grant agreement No.\
\href{https://doi.org/10.3030/101072410}{101072410}.
All content represents the opinion of the authors, which is not necessarily
shared or endorsed by their respective employers and/or sponsors.
\end{acks}

\section*{GenAI Usage Disclosure}
We used generative AI tools to assist with language polishing and with code cleaning and optimization during manuscript preparation. The authors carefully reviewed all generated content and take full responsibility for the final manuscript.

\bibliographystyle{ACM-Reference-Format}
\balance
\bibliography{references.bib}

\end{document}